\documentclass[11pt, oneside]{article}  
\usepackage{amsmath,amsfonts,amssymb,amsthm, mathtools,xfrac,bm}
\usepackage{graphicx,xcolor}
\usepackage[utf8]{inputenc}
\usepackage[english]{babel}
\usepackage[mathlines]{lineno}
\usepackage{listings}
\usepackage{textcomp}
\usepackage{appendix}
\usepackage{float}
\usepackage{comment}
\usepackage{stmaryrd}
\usepackage{siunitx}[=v2]
\usepackage{commath}
\usepackage[affil-it]{authblk}
\usepackage[outdir=./Figures/epstopdf/]{epstopdf}
\numberwithin{equation}{section}
\usepackage[]{siunitx}[=v2]
\usepackage[numbers]{natbib}
\usepackage[unicode, pdfpagelabels,bookmarks,hyperindex,hyperfigures]{hyperref}
\usepackage{cleveref}
\usepackage{array} 
\usepackage{caption}
\usepackage{subcaption}
\usepackage{booktabs}
\usepackage{diagbox}
\usepackage{array}
\usepackage{changepage}
\newcommand{\biot}{\text{Bi}}
\newcommand{\bound}{\partial \Omega}
\newcommand{\tpin}{T_{\text{PINN}}}

\graphicspath{{Figures/},{Figures/pdf/}}

\usepackage[acronym, symbols, nonumberlist, nopostdot, nogroupskip, toc]{glossaries-extra}
\setabbreviationstyle[acronym]{long-short}
\newglossary*{sub}{Subscripts}
\newglossary*{greek}{Greek}
\newglossary*{roman}{Roman}
\makeglossaries
\author[1]{Arijit Hazra \thanks{\,Corresponding author. Email: \texttt{ahazra@iitpkd.ac.in}}}
\author[2]{Prahar Sarkar \thanks{Email: \texttt{prahars.me.rs@jadavpuruniversity.in}}}
\author[2]{Sourav Sarkar \thanks{\,Corresponding author. Email: \texttt{souravsarkar.mech@jadavpuruniversity.in}}}
\affil[1]{\footnotesize Department of Mechanical Engineering, Indian Institute of Technology, Palakkad, India}
\affil[2]{\footnotesize Department of Mechanical Engineering, Jadavpur University, Kolkata, India}
\title{Physics-Informed Neural Networks for Estimating Convective Heat Transfer in Jet Impingement Cooling: A Comparison with Conjugate Heat Transfer Simulations }
\newacronym{chtc}{CHTC}{convective heat transfer coefficient}
\newacronym{pinn}{PINN}{physics-informed neural network}
\newacronym{cht}{CHT}{conjugate heat transfer}
\newacronym{cfd}{CFD}{computational fluid dynamics}
\newacronym{ihtp}{IHTP}{inverse heat transfer}
\newacronym{dl}{DL}{deep learning}
\newacronym{gpu}{GPU}{graphics processing unit}
\newacronym{pde}{PDE}{partial differential equations}

\newglossaryentry{biot}{
type=roman,
  name={$\biot$},
  description={Biot number}
}
\newglossaryentry{cp}{
  type=roman,
  name={$c_p$},
  description={specific heat [\si{\joule \per \kg \per \kelvin}]}
}
\newglossaryentry{hconv}{
  type=roman,
  name={$h$},
  description={convective heat transfer coefficient [\si{\watt \per \m \squared \per \kelvin}]}
}
\newglossaryentry{vel}{
  type=roman,
  name={$\mathbf{v}$},
  description={jet velocity [\si{\m \per \s}]}
}

\newglossaryentry{pressure}{
  type=roman,
  name={$p$},
  description={pressure [\si{\pascal}]}
}
\newglossaryentry{Re}{
  type=roman,
  name={Re},
  description={Reynolds number}
}
\newglossaryentry{gbf}{
  type=roman,
  name={$\mathbf{g}$},
  description={ gravitational acceleration [\si{\m \per \s \squared}]}
}
\newglossaryentry{enthal}{
  type=roman,
  name={$\hat{h}$},
  description={specific enthalpy [\si{\joule \per \kg}]}
}
\newglossaryentry{tempfluid}{
  type=roman,
  name=$T_f$,
  description={temperature within the fluid domain [\si{\kelvin}]}
}
\newglossaryentry{tempsolid}{
  type=roman,
  name=$T_s$,
  description={temperature within the solid domain [\si{\kelvin}]}
}

\newglossaryentry{tempinit}{
  type=roman,
  name=$T_0$,
  description={initial temperature of the solid domain [\si{\kelvin}]}
}
\newglossaryentry{tempnondim}{
  type=roman,
  name=$T$,
  description={nondimensional temperature in the solid domain [\si{\kelvin}]}
}
\newglossaryentry{tke}{
  type=roman,
  name={$k$},
  description={turbulent kinetic energy 
  [\si{\joule \per \kg}]
  }
}
\newglossaryentry{tkemeanvel}{
  type=roman,
  name={$G_k$},
  description={turbulence kinetic energy production due to mean velocity gradients [\si{\meter \squared\per\s \cubed}]}
}
\newglossaryentry{tkeb}{
  type = roman,
  name={$G_b$},
  description={turbulence kinetic energy production due to buoyancy [\si{\meter \squared\per\s \cubed}]}
}

\newglossaryentry{loss}{
  name={\ensuremath{\mathcal{L}}},
  description={Loss function used in the neural network training}
}
\newglossaryentry{dnozzle}{
type=roman,
  name={$D_n$},
  description={nozzle diameter [\si{\m}]}
}
\newglossaryentry{lnozzle}{
type=roman,
  name={$L_n$},
  description={nozzle length [\si{\m}]}
}
\newglossaryentry{hnp}{
type=roman,
  name={$H_n$},
  description={nozzle-to-plate-surface distance [\si{\m}]}
}

\newglossaryentry{rplate}{
type=roman,
  name={$R$},
  description={radius of the target plate [\si{\m}]}
}

\newglossaryentry{tplate}{
type=roman,
  name={$L$},
  description={thickness of the target plate [\si{\m}]}
}
\newglossaryentry{rho}{
  type = greek,
  name={$\rho$},
  description={density of the fluid [\si{\kg \per \metre\cubed}]}
}
\newglossaryentry{mu}{
  type = greek,
  name={$\mu$},
  description={dynamic viscosity [\si{\kg \per \m \per \s}]}
}
\newglossaryentry{kappa}{
  type = greek,
  name={$\kappa$},
  description={thermal conductivity [\si{\watt\per\meter\per\kelvin}]}
}
\newglossaryentry{tdr}{
  type = greek,
  name={$\varepsilon$},
  description={turbulent dissipation rate [\si{\meter \squared\per\s \cubed}]}
}
\newglossaryentry{alpha}{
  type = greek,
  name={$\alpha$},
  description={thermal diffusivity $\frac{\kappa}{\rho c_p}$ [\si{\m \squared \per \s}]}
}

\newglossaryentry{sub_f}{type=sub,
  name={\ensuremath{f}},
  description={fluid properties}}

\newglossaryentry{sub_s}{type=sub,
  name={\ensuremath{s}},
  description={solid properties}}
  
\newglossaryentry{eff}{type=sub,
  name={eff},
  description={effective}}
  
\newglossaryentry{avg}{type=sub,
  name={avg},
  description={spatially averaged}}
  
\newglossaryentry{fc}{type=sub,
  name={\ensuremath{fc}},
  description={forced convection }}
  
\newglossaryentry{fr}{type=sub,
  name={\ensuremath{fr}},
  description={free convection }}
  
\newglossaryentry{amb}{type=sub,
  name={\ensuremath{\infty}},
  description={ambient}}
  
\newglossaryentry{air}{type=sub,
  name={air},
  description={air properties}}

\date{}
\begin{document}
\sloppy
\maketitle
\begin{abstract}
Efficient cooling is vital for the performance and reliability of modern systems such as electronics, nuclear reactors, and industrial equipment. Jet impingement cooling is widely used for its high local heat transfer rates. Accurate estimation of \gls{chtc} is essential for design, simulation, and control of thermal systems. However, estimating spatially varying \glspl{chtc} from limited and noisy temperature data poses a challenging inverse problem. This study presents a \gls{pinn} framework to estimate both averaged and spatially varying \glspl{chtc} at the fluid–solid interface in a jet impingement setup at Reynolds number 5000. The model uses sparse and noisy temperature data from within the solid and embeds the transient heat conduction equation along with boundary and initial conditions into its loss function. This enables inference of unknown boundary parameters without explicit modeling of the fluid domain. Validation is performed using synthetic temperature data from high-fidelity \gls{cht} simulations. The framework is tested under various additive Gaussian noise levels (up to \SI{30}{\percent}) and sampling rates \SIrange{0.25}{4.0}{\per \second}. For noise levels up to \SI{10}{\percent} and sampling rates of \SI{0.5}{\per \second} or higher, estimated \glspl{chtc} match \gls{cht}-derived benchmarks with relative errors below \SI{7.6}{\percent}. Even under high-noise scenarios, the framework maintains predictive accuracy when time resolution is sufficient. These results highlight the method’s robustness to noise and sparse data, offering a scalable alternative to traditional inverse methods, experimental measurements, or full \gls{cht} modeling for estimating boundary thermal parameters in real-world cooling applications.

{\bf keywords:} Jet impingement cooling, Physics-informed neural network, Inverse Heat Transfer, Conjugate heat transfer
\end{abstract}
\addcontentsline{toc}{section}{Abstract}

\glsaddall
\printglossary[title= Nomenclature, type=roman, nonumberlist]
\printglossary[title= Greek symbols, type=greek, nonumberlist]
\printglossary[type=sub,title={List of Subscripts}]
\printglossary[type=\acronymtype,nonumberlist]

\section{Introduction} \label{sec:intro}
Efficient thermal management is critical for the reliability and performance of a wide range of modern engineering systems, including high-powered electronic devices \cite{ventola_rough_2014, mahajan_cooling_2006}, high-performance computing devices \cite{xue_high_2020},  aerospace propulsion components \cite{han_recent_2004}, spacecraft \cite{silk_spray_2008}, electric vehicle battery packs \cite{zhao_comprehensive_2024}, power plants and modern manufacturing \cite{sengupta_use_2005, rivallin_general_2001, hackenhaar_experimental-numerical_2020}. These systems often operate under high thermal loads where inadequate heat dissipation can lead to overheating, material degradation, or complete system failure \cite{heimerson_adaptive_2022}.

Jet impingement cooling \cite{zuckerman_jet_2006} is extensively used due to its efficient heat dissipation mechanism. This method involves localized heat transfer with a high-speed stream of coolant, which can be either air or liquid. While highly efficient, the method poses analytical challenges due to substantial convective fluxes at stagnation points and their spatial variability. In jet-impingement-based cooling systems, the \glsxtrfull{chtc} at the solid–fluid interface is a key parameter that characterizes the local intensity of heat transfer. Accurate knowledge of this parameter is essential for both thermal design and model validation. However, due to the complexity of the flow field, including strong turbulence, separation, and recirculation zones, it exhibits sharp spatial gradients that are often difficult to measure or predict accurately.

Conventional approaches to estimating \glsxtrshort{chtc} rely on either direct experimental measurements or high-fidelity simulations involving \glsxtrfull{cht}. On the experimental side, inferring \glsxtrshort{chtc} typically requires high-resolution measurements of surface temperatures and heat fluxes, commonly obtained using techniques such as embedded thermocouples, infrared thermography \cite{carlomagno_infrared_2010}, or heat flux sensors \cite{knauss_novel_2009}. While these methods can provide valuable insights, they often involve considerable instrumentation complexity, potential intrusiveness, and sensitivity to measurement uncertainties—particularly in transient regimes or when access to the measurement surface is limited. As a result, direct experimental approaches are less frequently adopted for resolving spatially distributed \glspl{chtc} in practical settings.

Numerical simulations using \glsxtrfull{cfd} coupled with \glsxtrshort{cht} modeling offer an alternative for estimating local heat fluxes on the solid-liquid interface \cite{zhu_-depth_2017, yang_numerical_2007}. However, these methods face serious limitations, including high computational cost due to the need to resolve complex flow features, dependence on effective turbulence models and boundary conditions that may introduce modeling uncertainties, and the requirement for complete knowledge of the fluid domain and material properties.

These challenges underscore the need for alternative approaches to estimating boundary parameters like \gls{chtc}, especially when direct measurements are impractical or impossible. \glsxtrfull{ihtp} methods \cite{ozisik_inverse_2021} provide a cost-effective and non-intrusive approach for inferring parameters from sparse and noisy internal temperatures, which are significantly easier to acquire. They eliminate the necessity for complex instrumentation or full \gls{cht} simulations and are flexible enough to handle both constant and spatially variable coefficients across diverse geometric and flow scenarios.

\glsxtrshort{ihtp} problems, particularly those involving the estimation of thermal boundary parameters such as \glspl{chtc}, are widely encountered in engineering domains, including power systems \cite{duda_new_2009}, aerospace applications \cite{nakamura_inverse_2014}, material processing \cite{luchesi_inverse_2012}, and metallurgical engineering \cite{wang_mold_2012}. These problems are inherently ill-posed, making them sensitive to stochastic noise and requiring sophisticated techniques like the sequential function specification method \cite{bauzin_3d-transient_2020}, Tikhonov regularization \cite{bozzoli_estimation_2014} or gradient-based optimization \cite{han_estimation_2019} to achieve stability. However, the need for repeated forward simulations and complex physical modeling often renders these methods computationally expensive and challenging to implement, especially in nonlinear or highly dynamic systems. Additionally, traditional approaches may struggle to accurately capture heat transfer behavior in scenarios involving coupled effects like radiation and convection, such as the cooling of high-temperature steel surfaces undergoing oxidation \cite{malinowski_inverse_2022}. 

Traditional inverse methods for heat transfer problems, such as those based on finite difference or finite element solvers coupled with optimization or regularization frameworks, suffer from several critical drawbacks. These include manual tuning of regularization parameters to stabilize solutions—often at the cost of reduced accuracy, as the inverse problems are highly sensitive to measurement noise. They are also computationally expensive, as they typically involve repeated evaluations of forward models and their derivatives. Additionally, these methods struggle to flexibly estimate spatially or temporally varying parameters, especially in nonlinear or discontinuous settings \cite{vogel_computational_2002}.

Over the last decade, the early attempts \cite{tamaddon-jahromi_data-driven_2020, czel_simultaneous_2014} to use \glsxtrfull{dl} to tackle scientific and engineering challenges have offered promising alternatives in solving \glsxtrshort{ihtp} problems by mitigating some of the aforementioned issues. \glsxtrshort{dl}-based techniques are highly effective in learning complex nonlinear relationships and generalizing from sparse or noisy data, with the added advantage of being highly customizable in terms of their architecture. Once these models are trained, they can deliver real-time predictions, making them highly suitable for cutting-edge applications. Their parallelizable nature further enhances computational efficiency through the use of \glsxtrfull{gpu} acceleration. Nonetheless, initial deep learning methodologies for solving inverse problems present certain drawbacks. These approaches rely solely on data, disregarding the underlying physical laws, potentially resulting in predictions that contravene established physical principles. To address this, one can generate high-quality data sets, albeit at the expense of substantial offline computational costs. Additionally, they perform poorly when confronted with sparse, noisy, or insufficient data sets.

\Glsxtrfull{pinn} \cite{lagaris_artificial_1998,raissi_physics-informed_2019, cuomo_scientific_2022} have revolutionized the field of scientific machine learning with a simple yet elegant approach. They approximate solution variables while embedding key components of the governing mathematical model—such as \glsxtrfull{pde}, boundary conditions, and initial conditions—directly into the learning process through a composite loss function optimized via gradient-based algorithms. This integration improves accuracy and robustness while significantly reducing dependence on large datasets. By enforcing physical constraints, \glspl{pinn} lower the risk of overfitting and enhance model interpretability. They also exhibit robustness to random noise, providing solutions that adhere to core physical principles—an essential benefit when addressing complex inverse problems. Additionally, a significant strength of PINNs is their unified framework: the same model can be used for both forward and inverse problems, with parameters to be estimated being introduced as extra trainable variables in case of inverse analysis. Moreover, an additional key benefit of PINNs is their flexibility, allowing straightforward integration of further physical constraints or pre-existing information into the framework.

In this study, we present a framework using \glspl{pinn} for the inverse estimation of \glspl{chtc} at the fluid–solid boundary in a jet cooling system. The main aims and contributions of our work are threefold. Firstly, we design a \gls{pinn} architecture incorporating the transient heat conduction equation alongside pertinent boundary conditions to estimate either constant or spatially variable \glspl{chtc} from limited temperature data collected at a few spatial locations in the solid domain. This approach allows for both averaged and spatially boundary parameter estimation, catering to different modeling accuracy requirements. While the proposed framework is generally applicable to a wide class of metal cooling problems, in this work we focus specifically on steel cooling under jet impingement, owing to its industrial relevance and well-characterized material properties. Secondly, we verify the precision of the estimated heat transfer coefficients by comparing them with high-fidelity \gls{cht} simulations, which act as the ground truth in this study. This validation confirms the credibility of our \gls{pinn}-based inverse methodology as a viable alternative to traditional numerical modeling techniques. Lastly, we perform a comprehensive robustness assessment by introducing synthetic noise and testing various data sampling rates for temperature readings and analyzing their impact on the convergence, prediction accuracy, and generalization capabilities of the trained \glspl{pinn}.

The rest of the paper is structured as follows. \Cref{sec:prob_form} covers the comprehensive fluid flow-heat transfer coupled problem, including governing equations, boundary conditions for conjugate heat transfer modeling, turbulence modeling, and the setup of the inverse problem. In \Cref{sec:pinns}, we outline the \gls{pinn} architecture, the loss function components, and the training strategy. \Cref{sec:implementation} offers an in-depth explanation of \gls{cht} simulations, which act as both validation benchmarks and experimental proxies, alongside the \glspl{pinn} implementations. Numerical results for estimating both constant and spatially varying heat transfer coefficients are presented in \Cref{sec:results}, coupled with a thorough comparison against \gls{cht} methods and analysis under various noise levels and sampling rates. Finally, \Cref{sec:conclusion} discusses our work's implications, limitations, and potential future developments.
\section{Mathematical formulation} \label{sec:prob_form}
\subsection{Governing equations for three-dimensional jet impingement cooling with conjugate heat transfer} \label{sec:cfd_mod}
\begin{figure}[h!]
	\centering
\includegraphics[width=1.0\textwidth]{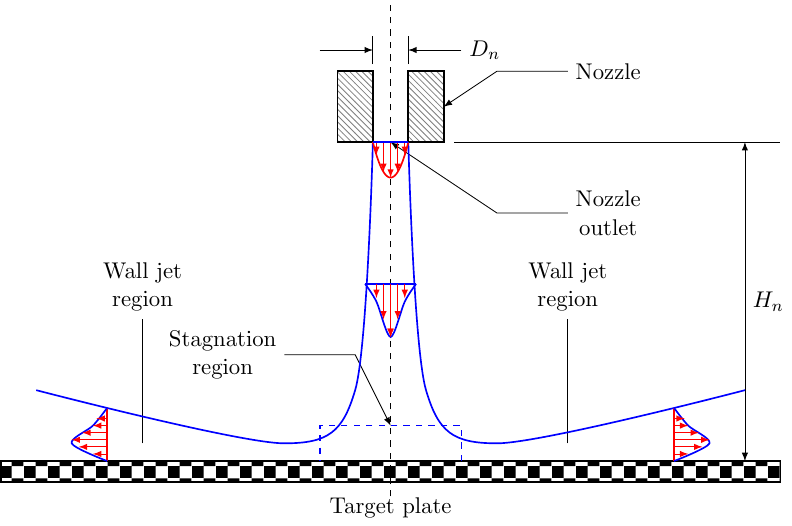}
\caption{Architecture of a Physics-Informed Neural Network (PINN) for parameter estimation. The symbol $\lambda$ represents the parameters to be estimated, which may include a constant convective heat transfer coefficient $h_{fc}$ or the coefficients of a polynomial approximation of $h_{fc}$. The notation $\partial_\xi^n$ denotes the $n$th-order partial derivative with respect to a generic variable $\xi$.}
\label{fig:jet-impingement-process}
\end{figure}
The fluid flow and heat transfer phenomena in the jet impingement cooling (\Cref{fig:jet-impingement-process}) problem are governed by the steady-state conservation equations for mass, momentum, and energy. For an incompressible flow, the equation of mass conservation is given by:
\begin{equation}
\nabla \cdot (\rho_f \mathbf{v}) = 0,
\end{equation}
where $\rho_f$ denotes the fluid density and \gls{vel} is the velocity vector.

The momentum equation in tensor form, incorporating the effects of turbulent viscosity, is given by \cite{pope_turbulent_2000}:
\begin{equation}
\rho_f (\mathbf{v} \cdot \nabla) \mathbf{v} = -\nabla p + \nabla \cdot \left[ \mu_{\text{eff}} \left( \nabla \mathbf{v} + (\nabla \mathbf{v})^T \right) \right] + \rho_f \mathbf{g}
\end{equation}
where \gls{pressure} is pressure, $\mu_{\text{eff}}$ is the effective dynamic viscosity, and $\rho_f\mathbf{g}$ represents the gravitational body force.

The energy equation for the fluid domain, expressed in terms of specific enthalpy \gls{enthal}, is:
\begin{equation}
\nabla \cdot (\rho_f \mathbf{v} \hat{h}) = \nabla \cdot (\kappa_f \nabla T_f),
\end{equation}
where \gls{tempfluid} denotes the temperature, and $\kappa_f$ represents the effective thermal conductivity of the jet fluid.

\paragraph{Turbulence modeling}

The turbulence in the impinging jet is resolved using the Renormalization Group (RNG) $k$--$\varepsilon$ model. This model introduces additional accuracy in rapidly strained and swirling flows, which are typical in impingement configurations. The transport equations for turbulent kinetic energy \gls{tke} and its dissipation rate \gls{tdr} are defined as follows:

\begin{equation}
\frac{\partial (\rho_f k)}{\partial t} + \nabla \cdot (\rho_f \mathbf{v} k) = \nabla \cdot \left( \alpha_k \mu_{\text{eff}} \nabla k \right) + G_k + G_b - \rho_f \varepsilon,
\end{equation}

\begin{equation}
\frac{\partial (\rho_f \varepsilon)}{\partial t} + \nabla \cdot (\rho_f \mathbf{v} \varepsilon) = \nabla \cdot \left( \alpha_\varepsilon \mu_{\text{eff}} \nabla \varepsilon \right) + C_{1\varepsilon} \frac{\varepsilon}{k}(G_k + C_3 G_b) - C_{2\varepsilon} \rho_f \frac{\varepsilon^2}{k},
\end{equation}

where \gls{tkemeanvel} is the turbulence kinetic energy production due to mean velocity gradients, \gls{tkeb} represents the buoyancy-induced production, and the constants are set as $C_{1\varepsilon} = 1.42$, $C_{2\varepsilon} = 1.68$, $\alpha_k = \alpha_\varepsilon = 1.393$, and $C_\mu = 0.0845$.

\paragraph{Conjugate heat transfer modeling.}
To account for heat conduction within the target plate, the transient heat conduction equation
\begin{equation}
\rho_s c_{p,s} \frac{\partial T_s}{\partial t} = \nabla \cdot (\kappa_s \nabla T_s),
\end{equation}
is solved in the solid domain with the convective boundary conditions 
     \begin{align}
      -\kappa_s\frac{\partial T_s}{\partial \eta} &=  h(T_s-T_\infty), \quad \mathbf{x} \in \partial \Omega \setminus \partial \Omega_i   \\
      -\kappa_s\frac{\partial T_s}{\partial \eta} &=  0, \quad \mathbf{x} \in \partial \Omega_i.    
     \end{align}
In this context, $T_s$ represents the temperature within the solid. The parameters $\rho_s$, $c_{p,s}$, and $\kappa_s$ denote the solid's density, specific heat capacity, and thermal conductivity, respectively. The bottom surface $\partial \Omega_b$ is insulated, while all other surfaces, except for $\partial \Omega_b$, termed $\partial\Omega_c = \partial \Omega \setminus \partial \Omega_b$, undergo convection. \gls{hconv} denotes the convective heat transfer coefficient at surfaces exposed to either ambient air or an impinging fluid. It varies spatially across different surface regions, reflecting localized thermal interactions. Forced convection occurs on the top surface due to jet impingement, while the vertical cylindrical surface experiences free (natural) convection driven by the surrounding ambient air.

Furthermore, we have a known initial condition, expressed as:
    \begin{align}
T_s(x,y,0) &= T_0, \quad \mathbf{x} \in \Omega
\end{align}

At the fluid–solid interface, conjugate heat transfer conditions are enforced by ensuring continuity of both temperature and heat flux normal to the surface:
\begin{align}
T_f &= T_s, \\
-\kappa_f \frac{\partial T_f}{\partial n} \bigg|_{\text{fluid}} &= -\kappa_s \frac{\partial T_s}{\partial n} \bigg|_{\text{solid}}.
\end{align}

\paragraph{Limitations of conjugate heat transfer modelling}
As briefly touched upon in \Cref{sec:intro}, integrating \glsxtrshort{cht} modeling with \glsxtrfull{cfd} provides detailed insights into metal cooling via jet impingement. However, these predictive methods necessitate extensive prior knowledge of flow conditions, including boundary conditions. Estimating \glsxtrshort{chtc} using \glsxtrshort{cfd} is not only computationally demanding but also highly sensitive to boundary parameters, turbulence models, and mesh quality. Moreover, in numerous practical cases, measurements are feasible only in the solid domain, making \glsxtrshort{cfd} simulations impractical when fluid domain characteristics are unknown. Given these limitations, we concentrate on inverse analysis using a physics-informed neural network as an alternative technique for estimating the convective heat transfer coefficient.

Despite the inherent challenges of \gls{cht} simulations in accurately predicting \glspl{chtc} in real-world scenarios—largely due to their dependence on fully known and precisely defined flow conditions—we use them in this study both as a benchmark and, with added Gaussian noise, as a surrogate for experimental data. This approach is justified by the controlled nature of the simulations, where the entire flow field is known. As a result, these limitations do not affect their validity for evaluating the accuracy of the \glspl{pinn}-based inverse estimation framework which is detailed in the following subsection.

\subsection{Inverse problem formulation}
\begin{figure}[!ht]
	\centering
\includegraphics[width=1.0\textwidth]{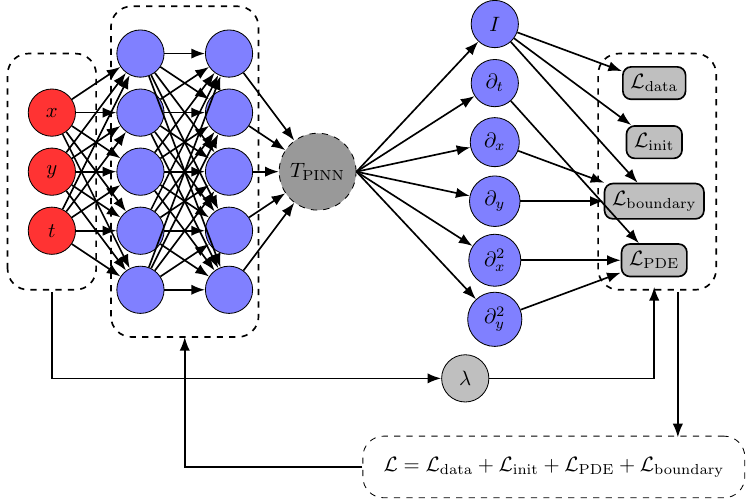}
\caption{
	\textbf{Left panel}: Schematic of the jet impingement cooling of a thin disc with radius $R$ and thickness $L$. 
	\textbf{Right panel}: Computational domain and boundary conditions used for inverse analysis. Owing to the axisymmetry of the centrally impinging jet, a representative 2D slice is extracted to enable estimation of the convective heat transfer coefficient over the entire domain. The magnified view on the right illustrates the applied boundary conditions: forced convection on the top surface (jet side), free convection on the right vertical wall, an adiabatic condition on the bottom wall, and a homogeneous Neumann condition on the left wall representing thermal symmetry.
}

\label{fig:modelprob}
\end{figure}

The central parameter estimation problem of this paper can be described as follows:
 \textit{Given a set of noisy, sparse temperature measurements $T_s(x_m,y_m,t^{(n)})$ located inside the solid domain at various points $x_m,~y_m$ (depicted as red dots in \Cref{fig:modelprob}) and measured at different times $t^{(n)}$, we need to estimate the key parameter: the convective coefficients $h_{fc}$ resulting from jet impingement where $h_{fc}$ might either be constant or spatially varying.}
 
We consider two scenarios: Initially, we examine the estimation of a constant forced convective coefficient on the top surface ($h_{fc}$); this fixed convective heat transfer coefficient can be seen as the spatial average of the convective heat transfer coefficient. Subsequently, we explore estimating a spatially variable heat transfer coefficient $h_{fc}(x)$, applied as a parametric function of $x$ along the top surface where forced convection occurs. The free convection coefficient $h_{fr}$ is presumed as known.
 
\paragraph{Simplified two-dimensional transient heat conduction formulation}
We analyze an axis-symmetric jet emanating centrally from a nozzle atop a metal plate, which allows us to simplify the problem using a two-dimensional plane (see \Cref{fig:modelprob}). By assuming the jet is perfectly symmetric relative to the heated metal, further simplifications are possible. We focus on a thin central slice of the disc illustrated in \Cref{fig:modelprob}. Consequently, the model problem reduces to 2D transient heat equations, with insulated boundaries on the left (due to thermal symmetry) and bottom, forced convection on the top, and free convection on the right.

Thus, our model heat conduction problem can be described by the following equation:
\begin{align}
    \frac{\partial T_s}{\partial t} &= \alpha_s  \left(\frac{\partial ^2T_s}{\partial x^2} + \frac{\partial ^2T_s}{\partial y^2}\right), \quad (x,y)  \in  \Omega,~~t \in [0,t_n]
    \label{eqn:heat_cond}
    \end{align}
    and the initial and boundary conditions are as follows:
    \begin{align}
\textnormal{IC:}\quad T_s(x,y,0) &= T_0, \quad (x,y) \in \Omega \notag \\
\textnormal{BC:} \quad 
    h_{fc}(T_s-T_\infty) & = -\kappa_s\frac{\partial T_s}{\partial y}(x, L,t) \notag \\
    h_{fr}(T_s-T_\infty) &= -\kappa_s\frac{\partial T_s}{\partial x}(R,y,t) \notag \\
    -\kappa_s\frac{\partial T_s}{\partial y}(x,0,t)&=0, \quad 
    -\kappa_s\frac{\partial T_s}{\partial x}(0,y,t)=0, \notag 
\end{align}
where \gls{rplate} and \gls{tplate} are the radius and thickness of the circular plate respectively.

\paragraph{Nondimensionalized form of the equation.}
For subsequent analysis, we reformulate the equations in a nondimensionalized form, as this approach is known to enhance the training capabilities and accuracy of PINNs \cite{kissas_machine_2020}. We transform all the primary variables as follows:
\begin{align*}
\tilde{T} = \frac{T_s-T_\infty}{T_0-T_\infty}, ~~\tilde{x} = \frac{x}{R}, ~~\tilde{y} = \frac{y}{R}, ~~\tilde{t} = \frac{\alpha_s t}{R^2}
\end{align*}
We non-dimensionalize the spatial variables using the characteristic length \gls{rplate}, justified by the fact that the plate thickness is much smaller than its radius, i.e. $\dfrac{L}{R}<<1$.

To maintain clarity in exposition, we omit the tildes henceforth and use the same notation without the tildes even for the nondimensional variables throughout the rest of the paper.

Using the above scaling and notation, the nondimensionalized form of the heat conduction equation becomes:
\begin{align}
    \frac{\partial T}{\partial t} - \left(\frac{\partial ^2 T}{\partial x^2} + \frac{\partial ^2 T}{\partial y^2}\right)=0,  \quad (x,y)\in [0,1]\times[0,y_r],~t \in [0,t_f]
    \label{eqn:nond_heat_cond}
\end{align}
where $y_r = \dfrac{L}{R},~t_f = \dfrac{\alpha_s t_n}{R^2}$. The initial and boundary conditions are specified as:
\begin{align}
    T(x,y,0) &= T_0(x,y) = 1 \notag \\
    \dpd{T}{y}(x,y_r,t)+ \biot_{fc}~T &= 0, \quad \dpd{T}{y}(x,0,t) = 0 \notag\\
    \dpd{T}{x}(1,y,t) + \biot_{fr}~T &= 0, \quad \dpd{T}{x}(0,y,t) = 0 \label{eqn:heat_non_dim_aux_con}
\end{align}

All subsequent equations are expressed in terms of nondimensional variables. The corresponding dimensional quantities can be recovered by inverting the applied scaling transformations.
Since the analysis is conducted using a nondimensionalized formulation, the inverse problem is also expressed in terms of nondimensional Biot numbers——$\biot_{fc}$ representing forced convection on the upper surface, and $\biot_{fr}$ representing free convection along the right edge.

\section{PINNs for temperature field reconstruction} \label{sec:pinns}
In our study, we employ \Glspl{pinn}\cite{raissi_physics-informed_2019, he_physics-informed_2021} to approximate the temperature field \( T(x,y,t) \) over the 2D domain. The PINN is a fully-connected deep neural network that takes as input the spatial coordinates and time, i.e., \( (x,y,t) \), and outputs the predicted temperature \(  T(x,y,t) \approx T_{\text{PINN}}(x,y,t) \). 

\begin{figure}[h!]
	\centering
\includegraphics[width=1.0\textwidth]{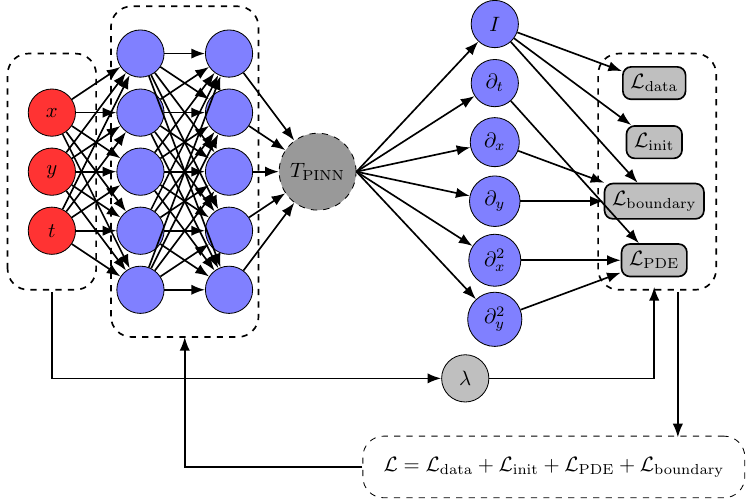}
\caption{Architecture of a Physics-Informed Neural Network (PINN) for parameter estimation. The symbol $\lambda$ represents the parameters to be estimated, which may include a constant convective heat transfer coefficient $h_{fc}$ or the coefficients of a polynomial approximation of $h_{fc}$. The notation $\partial_\xi^n$ denotes the $n$th-order partial derivative with respect to a generic variable $\xi$.}
\label{fig:pinns}
\end{figure}

The network, as depicted, \Cref{fig:pinns} is designed with several hidden layers and nonlinear activation functions, allowing it to capture the complex spatiotemporal behavior of the cooling process.

The training of the PINN is driven by a composite loss function, defined as follows:
\begin{align}
    \mathcal{L} = \mathcal{L}_{\text{data}} + \mathcal{L}_{\text{PDE}} + \mathcal{L}_{\text{boundary}} +  \mathcal{L}_{\text{initial}}
    \label{eq:loss_func}
\end{align}
where:
\begin{itemize}
    \item \textbf{Data loss (\( \mathcal{L}_{\text{data}} \))}: This term measures the discrepancy between the PINN-predicted temperature and the noisy synthetic data $T_{\text{meas}}$ obtained from \gls{cht} simulations. It is defined as:
    \[
    \mathcal{L}_{\text{data}} = \frac{1}{N_d} \sum_{i=1}^{N_d} \left( T_{\text{PINN}}(x_i,y_i,t_i) - T_{\text{meas}}(x_i,y_i,t_i) \right)^2,
    \]
    where \( N_d\) is the number of sensor data points.
    
    \item \textbf{PDE loss (\( \mathcal{L}_{\text{PDE}} \))}: This term enforces the governing 2D heat conduction equation:
    \[
    \frac{\partial T_{\text{PINN}}}{\partial t} =  \left( \frac{\partial^2 T_{\text{PINN}}}{\partial x^2} + \frac{\partial^2 T_{\text{PINN}}}{\partial y^2} \right).
    \]
    It is implemented as
    \[
    \mathcal{L}_{\text{PDE}} = \frac{1}{N_p} \sum_{i=1}^{N_p} \left( \frac{\partial T_{\text{PINN}}}{\partial t}(x_i,y_i,t_i) - \left[ \frac{\partial^2 T_{\text{PINN}}}{\partial x^2}(x_i,y_i,t_i) + \frac{\partial^2 T_{\text{PINN}}}{\partial y^2}(x_i,y_i,t_i) \right] \right)^2,
    \]
    where \( N_p\) represents the number of collocation points used for enforcing the PDE.
    
    \item \textbf{Boundary loss (\( \mathcal{L}_{\text{boundary}} \))}: To ensure that the boundary conditions are satisfied along the boundary \( \bound \) we apply a boundary loss function. Our boundary loss functions has four components for four walls and four distinct boundary conditions on those four walls of the domain and read as
    \begin{align*}
        \mathcal{L}_{\text{boundary}} &= \mathcal{L}_{\text{lb}} + \mathcal{L}_{\text{bb}} + \mathcal{L}_{\text{rb}} +
        \mathcal{L}_{\text{ub}} \\ 
        \mathcal{L}_{\text{lb}} &=\frac{1}{N_{lb}} \sum_{i=1}^{N_{lb}} \left(\frac{{\partial T_{\text{PINN}}}}{{\partial x}}(0,y_i,t_i)\big) \right)^2 \\
        \mathcal{L}_{\text{bb}} &=\frac{1}{N_{bb}} \sum_{i=1}^{N_{bb}} \left(\frac{{\partial T_{\text{PINN}}}}{{\partial y}}(x_i,0,t_i)\big) \right)^2 \\
        \mathcal{L}_{\text{rb}} &=\frac{1}{N_{rb}} \sum_{i=1}^{N_{rb}} \left(\frac{{\partial T_{\text{PINN}}}}{{\partial y}}(1.0, y_i,t_i) + \biot_{fr}~T_{\text{PINN}}\big) \right)^2\\
        \mathcal{L}_{\text{ub}} &=\frac{1}{N_{ub}} \sum_{i=1}^{N_{ub}} \left(\frac{{\partial T_{\text{PINN}}}}{{\partial y}}(x_i, y_r,t_i) + \biot_{fc}(x_i)~T_{\text{PINN}}\big) \right)^2
    \end{align*}
    with $N_{lb},~N_{bb},~N_{rb},~N_{ub}$ being the number of collocation points at left, bottom, right and uppper boundary respectively.

    \item \textbf{Initial loss (\( \mathcal{L}_{\text{initial}} \))}: To ensure that the initial conditions are satisfied over the interior of the computational domain \( \Omega\) and the follwoing term is added:
    \[
    \mathcal{L}_{\text{initial}} = = \frac{1}{N_i} \sum_{k=1}^{N_i} \left( T_{\text{PINN}}(x_k,y_k,0) - T_0(x_k,y_k)\right)^2,
    \]
    with \( N_i \) being the number of initial points.
\end{itemize}

 Furthermore, we add a $L^2$-regularization term that results in a smooth evolution in the loss functions and proper convergence.

Please note that we apply our method on the non-dimensionalized form of the PDE, boundary and initial conditions. This helps us in avoiding the data-sclaing that is generally employed for any data driven method. Instead of data scaling we employ physical scales to normalize our data.

\paragraph{PINNs for function estimation.}
To go beyond the assumption of a constant \glsxtrfull{chtc}—representing a spatially averaged value—and to capture its spatial variation, the inverse problem is reformulated as a parametric estimation task. In this approach, the spatially varying \gls{chtc} is expressed through a selected functional form, and the goal becomes inferring the optimal parameters that best represent this distribution.

By integrating the loss components as described in $\Cref{eq:loss_func}$, the PINN effectively learns the temperature field $T(x,y,t)$ while facilitating the robust estimation of both the spatially varying heat transfer coefficient $h_{fc}(x)$ and the constant $h_{fc}$.

\section{Data generation and numerical implementation}\label{sec:implementation}
To evaluate the accuracy and robustness of the \glspl{pinn} framework for the estimation of \glsxtrfull{chtc} using inverse analysis, we generate a synthetic data set using high-fidelity \glsxtrfull{cht} simulations of a jet impingement. The geometry and flow configurations used in the \gls{cht} simulations closely adheres to those described in the experimental study by Guo et al. \cite{guo_experimental_2017}, which investigated jet impingement cooling using embedded thermocouples.
The present simulation includes essential parameters like nozzle diameter and jet-to-target distance from their study to ensure the generated data set stays within an experimentally reliable range.

\subsection{Conjugate heat transfer simulation setup}

\paragraph{Geometry and boundary conditions.}
The geometry for the jet impingement simulation consists of a circular air jet impinging perpendicularly on a flat solid circular disc. The air jet is modeled as issuing from a nozzle of length $L_n = \SI{60}{\milli\meter}$ and inner diameter $D_n = \SI{6}{\milli\meter}$. The nozzle is positioned at a distance $H_n$ from the disc, with $H_n$ is set at a distance of  $4D_n$. The computational domain includes both the fluid domain and the solid plate for \gls{cht} simulations and computational domain covers region from the inlet of the nozzle up to the bottom of the circular disc The solid disc has dimensions of \SI{120}{\milli\metre} in breadth and \SI{15}{\milli\meter} in thickness, and is assumed to be made of stainless steel with temperature-independent thermal properties.

Boundary conditions include a constant velocity at the nozzle inlet corresponding to a Reynolds numbers of \num{5000}, a fixed temperature (\SI{293}{\kelvin}) for the incoming air, and a pressure outlet at the far boundary. The top and side boundaries are treated as symmetry or far-field with convective outflow conditions to minimize reflection. No-slip and no-penetration conditions are applied to all solid surfaces. The fluid–solid interface enforces continuity of temperature and heat flux. Given the chosen Reynolds number, the air is modeled as an incompressible ideal gas with temperature-independent properties, including a thermal conductivity of \SI{2.514e-02}{\watt\per\metre\per\kelvin} and a Prandtl number of \num{0.7309}. 
\begin{figure}[!ht]
	\centering
\includegraphics[width=1.1\textwidth]{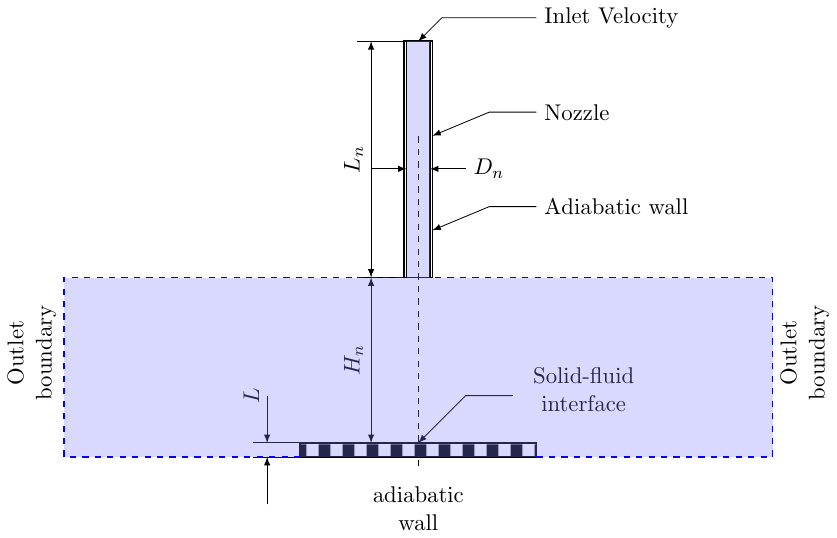}
\caption{Schematic of the front view of the computational domain used for 3D coupled \gls{cht} simulations of jet impingement cooling. The blue region depicts the fluid domain, and the rectangular region at the bottom with a checkerboard pattern represents the front view of the heated circular disc. The domain includes the jet nozzle, the surrounding air region, and the solid circular disc to fully capture conjugate heat transfer effects. Geometrical features such as the nozzle diameter $D_n$, nozzle length $L_n$, nozzle-to-plate distance $H_n$, and disc thickness $L$ are labeled. The dashed blue line represents the outlet boundary condition. The schematic is not drawn to scale and is intended to illustrate the physical layout, boundary conditions, and fluid–solid coupling.}
\label{fig:jet-impingement-CFD}
\end{figure}

\begin{table}[!ht]
\centering
\caption{Jet impingement nozzle and fluid parameters}
\begin{tabular}{llc}
\hline
\textbf{Parameter} & \textbf{Symbol} & \textbf{Value} \\
\hline
Nozzle length & \( L_n\) & \SI{60}{\milli\meter} \\
Nozzle inner diameter & \( D_n\) & \SI{6}{\milli\meter} \\
Reynolds number & \( \text{Re} \) & \num{5000} \\
Maximum Jet velocity & \( U_\text{max} \) & \SI{12.63}{\meter\per\second} \\
Nozzle-to-plate distance & \( H_n \) & $4D_n$ \\
Working fluid temperature & \( T_{\text{air}} \) & \SI{293}{\kelvin} \\
Thermal conductivity of air & \( \kappa_{\text{air}} \) & \SI{2.514e-02}{\watt\per\meter\per\kelvin} \\
Kinematic viscosity of air & \( \nu_{\text{air}} \) & \SI{1.516e-5}{\metre \squared \per \second} \\
Thermal diffusivity of air & \( \alpha_{\text{air}} \) & \SI{2.074e-5}{\metre \squared \per \second} \\
Prandtl number of air & \( \text{Pr}_{\text{air}} \) & 0.7309 \\
\hline
\end{tabular}
\label{tab:jet_params}
\end{table}

\paragraph{Solver and parameters.}
The simulations are carried out using transient solvers for incompressible flow in ANSYS Fluent.
The numerical solution is obtained using the finite volume method. The pressure–velocity coupling is resolved through the SIMPLE (Semi-Implicit Method for Pressure-Linked Equations) algorithm. Second-order upwind discretization is applied to the momentum and energy equations to ensure higher accuracy in the presence of strong convective gradients. Gradient calculations are performed using the least-squares cell-based method to improve accuracy near curved and irregular boundaries.

A structured, axisymmetric mesh is generated with local refinement near the stagnation zone and the fluid–solid interface to capture steep thermal and velocity gradients. The inlet boundary condition specifies a uniform velocity and constant temperature corresponding to the jet flow. The outlet is assigned a zero-gauge pressure condition. The bottom of the steel plate is modeled with either a fixed temperature or adiabatic condition, depending on the cooling scenario. 
Grid independence is ensured via a grid refinement study near the jet axis and stagnation zone (see \Cref{app:gridind}).

\begin{table}[h!]
\centering
\caption{Simulation Parameters for Jet Impingement on Heated Plate}
\begin{tabular}{|l|p{8cm}|}
\hline
\textbf{Parameter} & \textbf{Value} \\
\hline
Viscous Model & RNG \( k\text{-}\varepsilon \) turbulence model \\
\hline
Material & \textbf{Fluid:} Air \newline 
\textbf{Plate:} Stainless Steel 304 \newline 
Density $\rho_s$: \SI{7920}{\kg \per \m \cubed} \newline 
Specific Heat $c_{p,s}$: \SI{538}{\joule \per \kg \per \kelvin} \newline 
Thermal Conductivity $\alpha_s$: \SI{16.27} {\watt \per \m \per \kelvin} \\
\hline
Initial Plate Temperature $T_0$ & \SI{673}{\kelvin} \\
\hline
Solution Method & SIMPLE algorithm with second-order discretization \\
\hline
Time Step & Transient simulation with time step size of  \SI{1.0e-04}{\s} (from literature), total duration: \SI{300}{\s} \\
\hline
\end{tabular}
\label{tab:sim_params}
\end{table}

\paragraph{Convergence and post-processing.}

The convergence of the simulation is ensured by monitoring residuals of all governing equations. Residuals for momentum and turbulence variables are set to a tolerance of $10^{-4}$, while the energy equation is solved until residuals fall below $10^{-6}$. Post-processing focuses on evaluating wall temperature distribution, local and area-averaged Nusselt numbers, velocity and turbulence profiles within the jet core and wall jet regions, and thermal penetration depth within the steel plate.

\paragraph{Data acquisition from synthetic measurement}
We choose exactly \num{8} thermocouples positioned \SI{0.9375}{\milli \metre} beneath the surface and situated at \SIlist{3.75; 11.25; 18.75; 26.25; 33.75; 41.50; 49.0; 56.5}{\milli \metre} from the left as depicted in \Cref{fig:modelprob}. The framework undergoes evaluation with varying levels of additive Gaussian noise ranging from \SIrange{0}{30}{\percent} and a sampling rate of \SIlist{0.25; 0.5; 1.0; 2.0; 4.0}{\per \s}.

\subsection{PINNs implementation} \label{subsec:pinn_imple}

We implement our \gls{pinn} model using the DeepXDE library \cite{lu_deepxde_2021} with a TensorFlow backend \cite{abadi_tensorflow_2016}. DeepXDE is specifically designed for solving problems involving physics-informed machine learning, providing a unified framework that supports multiple machine learning and deep learning libraries, such as TensorFlow and PyTorch, as backends. Additionally, it offers syntactic simplicity, making the implementation of \glspl{pinn} more accessible and efficient. 
In this subsection, we discuss the design of the neural network architecture, composition of loss function, training procedures, and validation strategies to ensure accurate and physically consistent predictions.
\paragraph{Data preprocessing.}
In \gls{dl}-based methods, data scaling is crucial for effective training. In our approach, we build the model using the nondimensionalized form of the governing equation \ref{eqn:nond_heat_cond} so that the resulting temperature data lies within the interval \numrange{0}{1}. We also normalize the simulated experimental data to ensure consistent scaling. Consequently, by nondimensionalizing the governing equations and boundary conditions, we eliminate the need for conventional scaling methods typically employed in \gls{dl}-based techniques.
\paragraph{Network architecture.}
\begin{itemize}
    \item Feedforward neural networks ($\tpin$) are employed to approximate $T(x,y,t) \approx \tpin$.
    \item The network accepts $(x,y,t)$ as input features to model the temperature field's spatiotemporal dynamics.
    \item \numrange{5}{6} hidden layers, each with \num{40} neurons, using $\tanh$ activation, are utilized to model partial differential equations.
    \item An inbuilt module of deepxde enforces initial conditions.
    \item The network's output predicts the temperature $T(x,y,t)$ while also calculating non-dimensional parameters, such as constant $\biot_{fc}$ (for estimation of spatially averaged $h_{fc}$) and polynomial coefficients that parametrize the $\biot_{fc}$ (for estimation of spatially varying $h_{fc}$). These parameters are trained as external variables.
    \item \textbf{Learning rate.} We use a learning rate of \numrange{1.0e-04}{3.0e-04}.
    \item \textbf{Optimizer.} Optimization is performed using the \textit{Adam} optimizer.
    \item \textbf{Sampling points.} The boundary points are randomly selcted from a range of \numrange{1000}{1500}. The initial condition sampling points are selected as half of the boundary points, while the training and testing domain points are set at \num{5} and \num{7} times the boundary points, respectively.
\end{itemize}

\paragraph{Hyperparameter tuning.}
Our experiments involve optimizing critical hyperparameters, including network architecture parameters (such as the number of hidden layers and neurons, and sampling points both at boundaries and within the domain. This is achieved through Bayesian optimization \cite{snoek2012practical} using the Optuna library \cite{akiba2019optuna}, augmented with early stopping via the median pruner.

\paragraph{Computational framework.}
The simulations in this study were performed on a high-performance workstation equipped with an Intel(R) Xeon(R) Silver 4214R 24-core processor, running at a base clock of 2.40 GB (boost up to 3.50 GB ), 64 GB of DDR4 RAM, and an NVIDIA RTX A4000 GPU with 16 GB of dedicated VRAM.

The system ran on a Linux-based operating system (Debian GNU/Linux 12 (bookworm)) and utilized Python 3.9.21 with Tensorflow 2.17.1 and the DeepXDE library 1.13.2 for implementing and training the \glspl{pinn}. With the computational resources outlined above, all inverse simulations were performed efficiently. The availability of GPU acceleration significantly reduced training times for the deep neural networks, with each test case—run over \num{50,000} iterations (epochs)—completing in approximately \SI{6}{\minute}.

\subsection{Error metric for inverse estimation} \label{subsec:err_metric}

To quantitatively assess the accuracy of the inverse estimation of the convective heat transfer coefficient, we compute the relative \( L^2 \) error between the estimated profile \( h_{\text{PINN}}(x) \) and the reference solution \( h_{\text{ref}}(x) \) obtained from \glsxtrshort{cht} simulations. The relative error is defined as
\begin{align}
\text{Relative Error} = \frac{\| h_{\text{PINN}}(x) - h_{\text{ref}}(x) \|_2}{\| h_{\text{ref}}(x) \|_2},
\label{eqn:error_profile}
\end{align}
where \( \| \cdot \|_2 \) denotes the discrete \( L^2 \)-norm over the evaluation domain. 

For the special case of estimating a spatially constant parameter \( h_{fc} \), the relative error simplifies to the absolute difference between the estimated and true values normalized by the magnitude of the true value, i.e.,
\begin{align}
\text{Relative Error} = \frac{ \left| h_{\text{PINN}} - h_{\text{ref}} \right| }{ \left| h_{\text{ref}} \right| }.
\label{eqn:error_const}
\end{align}

\section{Results and discussion} \label{sec:results}
We now present the results of the inverse estimation of \glsxtrfull{chtc}s using the proposed \glspl{pinn} framework. Two test cases are considered: a simplified case assuming a spatially constant  $h_{\text{avg}}$ (representative of spatially averaged convective heat transfer coefficient $h_{fc}$), and a more realistic case involving spatially varying $h_{fc}(x)$. For each scenario, we test robustness of our framework for boundary parameter estimation by comparing them against ground truth values extracted from \gls{cht} simulations. The results are evaluated in terms of error metrics, qualitative agreement, and sensitivity to noise and sampling rate.

\subsection{Estimation of spatially averaged convective heat transfer coefficient} \label{subsec:avg_estim}

In the first phase of our study, we assume a spatially uniform \gls{chtc} across the entire impingement surface. This simplification is commonly employed in system-level or lumped-parameter thermal models. Despite its simplicity, the case provides a suitable benchmark to test the numerical stability of the \glspl{pinn} framework and to evaluate its sensitivity to noise and sampling rate. The estimated $h_{\text{avg}}$ is compared against a spatially averaged value ($h_{\text{ref}}$) obtained from \gls{cht} simulations through numerical integration using
Simpson’s rule.

The PINN is implemented using a fully connected feedforward neural network comprising \num{5} layers. Following earlier hyperparameter optimization, the count of boundary points is determined to be \num{1115}. For training, the number of interior domain points is configured to be $5 \times$ and $7 \times$ the boundary points for testing, respectively, as detailed in \Cref{subsec:pinn_imple} alongside other hyperparameters.

\begin{figure}[htp!]
    \centering
\begin{adjustwidth}{-3cm}{0cm}
\begin{tabular}{l l}
\includegraphics[width=0.75\textwidth]{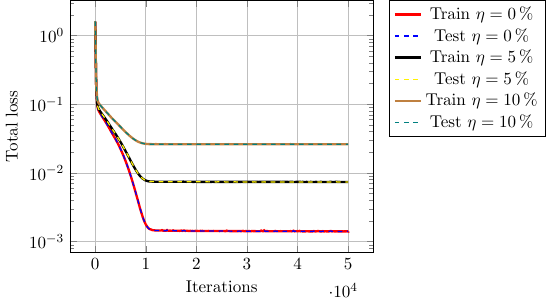} & \includegraphics[width=0.50\textwidth]{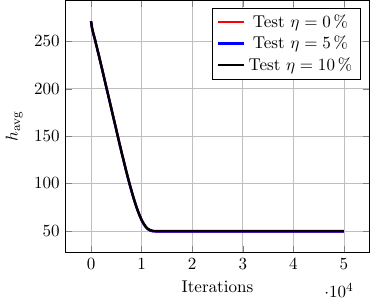} \\ 
\includegraphics[width=0.75\textwidth]{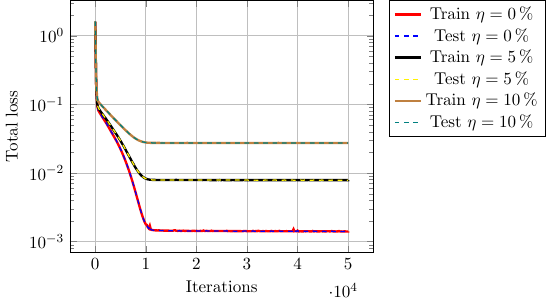} & \includegraphics[width=0.50\textwidth]{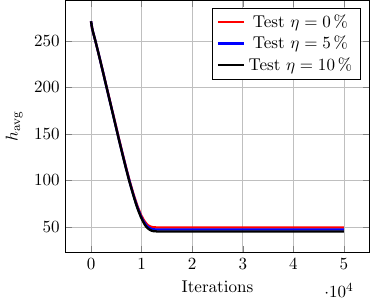} \\ 
\includegraphics[width=0.75\textwidth]{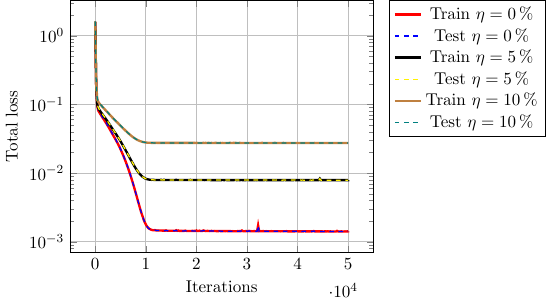} & \includegraphics[width=0.50\textwidth]{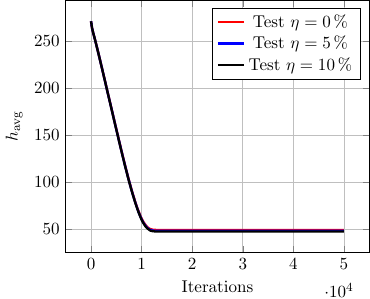}
\end{tabular}
\end{adjustwidth}
\caption{Convergence plots of total loss for both training and test data (left) and estimated $h_{\text{avg}}$ (right) on the upper surface of the disc for additive Gaussian noise \SIlist{0; 5; 10}{\percent} at different sampling rates.  Top. \SI{0.25}{\per \s}. Middle \SI{0.50}{\per \s}. Bottom. \SI{1.0}{\per \s}.}
\label{fig:h_loss_conv_avg1}
\end{figure}

\begin{figure}[htp!]
    \centering
    \begin{adjustwidth}{-3cm}{0cm}
 \begin{tabular}{l l}
\includegraphics[width=0.75\textwidth]{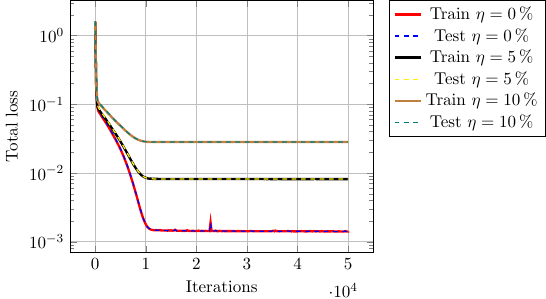} & \includegraphics[width=0.50\textwidth]{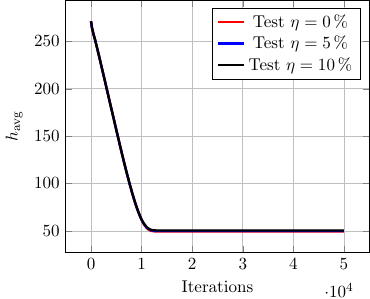} \\ 
\includegraphics[width=0.75\textwidth]{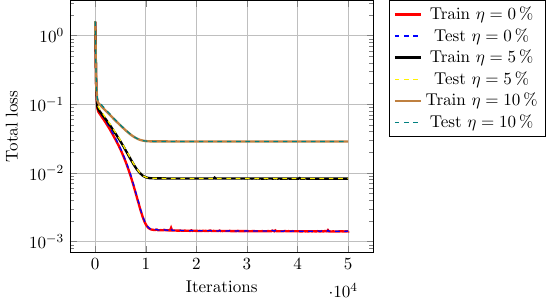} & \includegraphics[width=0.50\textwidth]{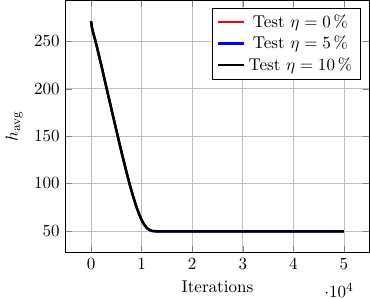}
\end{tabular}
\end{adjustwidth}
\caption{Convergence plots of total loss for both training and test data (left) and estimated $h_{\text{avg}}(x)$ (right) on the upper surface of the disc for additive Gaussian noise \SIlist{0; 5; 10}{\percent} at different sampling rates.  Top. \SI{2.0}{\per \s}. Bottom. \SI{4.0}{\per \s}.}
\label{fig:h_loss_conv_avg2}
\end{figure}

\Cref{fig:h_loss_conv_avg1,fig:h_loss_conv_avg2} shows the convergence of the loss function and estimated $h_{\text{avg}}$ over training iterations. The left panel illustrates the loss function decay, while the right panel shows the convergence behavior of the estimated heat transfer coefficient. We depict the convergence plots for three noise levels—low, medium, and high—corresponding to \SIlist{0;5;10}{\percent} additive Gaussian noise, respectively. For each noise level, results are obtained across five sampling rates: \SIlist{0.25; 0.5; 1.0; 2.0; 4.0}{\per \second}. 

\Cref{tab:noise_sr_error_avg_8_sensors} provides a quantitative assessment of the relative error (as defined in \Cref{subsec:err_metric}) for each configuration resulting from the additive Gaussian noise range of \SIrange{0}{30}{\percent} and sampling rate between \SIrange{0.25}{4.0}{\per \second}.

\begin{table}[H]
    \centering
    \begin{tabular}{c|rrrrr}
\toprule
\diagbox[width=4.5cm]{Noise (\si{\percent})}{Sampling rate}& 0.25 & 0.50 & 1.00 & 2.00 & 4.00 \\
\midrule
0.0 & 2.805 & 2.805 & 2.805 & 2.805 & 2.805 \\
2.0 & 2.241 & 1.111 & 2.241 & 3.370 & 2.805 \\
4.0 & 2.805 & 0.584 & 1.111 & 3.370 & 2.805 \\
5.0 & 2.805 & 1.714 & 1.111 & 3.935 & 2.805 \\
6.0 & 3.370 & 2.843 & 0.546 & 4.500 & 2.805 \\
8.0 & 3.370 & 3.408 & 0.019 & 4.500 & 3.370 \\
10.0 & 3.935 & 6.232 & 0.584 & 5.065 & 3.370 \\
15.0 & 4.500 & 10.186 & 2.278 & 6.759 & 3.935 \\
20.0 & 5.630 & 14.705 & 3.973 & 7.889 & 4.500 \\
25.0 & 6.195 & 18.659 & 5.668 & 9.019 & 5.065 \\
30.0 & 6.759 & 22.614 & 7.362 & 10.149 & 5.065 \\
\bottomrule
\end{tabular}
\caption{Relative error (defined in \Cref{eqn:error_const}) in the estimated averaged convective heat transfer coefficient $h_{\text{avg}}$ using \glspl{pinn}, with spatially averaged value $h_{\text{ref}}$ obtained from \gls{cht} simulations through
numerical integration taken as refernce, for various levels of additive Gaussian noise and sampling rate.}
    \label{tab:noise_sr_error_avg_8_sensors}
    \end{table}

Our results indicate that under low-noise conditions, sparsely sampled data perform comparably to high-frequency data. However, in high-noise scenarios, higher-resolution measurements demonstrate significantly greater robustness and reliability.

Interestingly, we observe that the estimation error does not always increase monotonically with noise level. In some cases, moderate noise leads to improved generalization, possibly due to regularizing effects in the training loss. This behavior, while counterintuitive, is consistent with similar observations in inverse analysis \cite{fernandez-martinez_effect_2014-1}.

\subsection{Estimation of spatially varying convective heat transfer coefficient} \label{subsec:spatially_variable}

We now consider the more challenging inverse problem of estimating a spatially varying convective heat transfer coefficient $h_{fc}(x)$ along the fluid–solid interface for jet-impingement cooling. For this study, the functional form of $h_{fc}(x)$ is assumed to follow a second-degree polynomial, consistent with expected heat transfer profiles in axisymmetric jet impingement configurations \cite{guo_experimental_2017}. The inverse problem is thus posed as a three-parameter estimation problem, with the aim of recovering the detailed spatial distribution of local heat transfer coefficients using only temperature measurements within the solid domain.

The PINN model used here is a feedforward neural network with \num{6} hidden layers, trained on \num{1375} boundary points, selected based on prior
hyperparameter tuning. All other configurations—such as the choice of optimizer, domain discretization, and loss formulation—remain similar as the previous test case.

\begin{figure}[!ht]
    \centering
    \begin{adjustwidth}{-1cm}{0cm}
    \begin{tabular}{l l}
\includegraphics[width=0.50\textwidth]{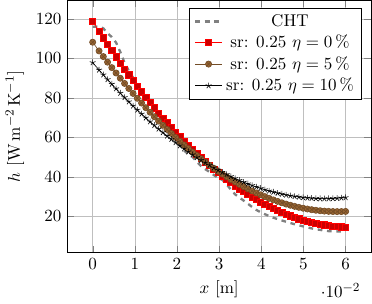} & \includegraphics[width=0.50\textwidth]{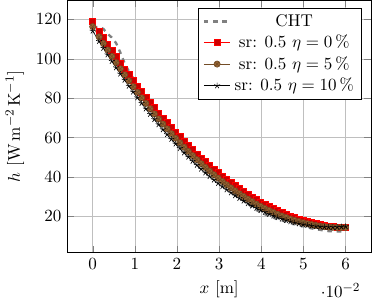} \\ 
\includegraphics[width=0.50\textwidth]{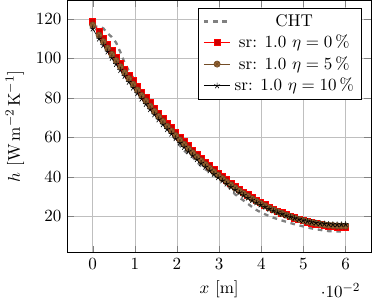} & \includegraphics[width=0.50\textwidth]{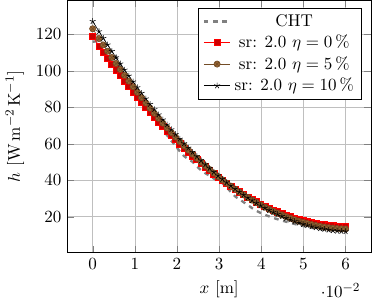}\\
\includegraphics[width=0.50\textwidth]{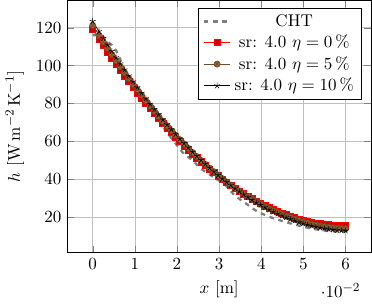} & \\
\end{tabular}
\end{adjustwidth}
    \caption{Estimated profiles of the convective heat transfer coefficient $h_{fc}(x)$ on the upper surface of the disc for additive Gaussian noise \SIlist{0; 5; 10}{\percent} at different sampling rates(sr).  Top Left. \SI{0.25}{\per \s} . Top Right. \SI{0.50}{\per \s}. Middle left. \SI{1.0}{\per \s}. Middle right. \SI{2.0}{\per \s}. Bottom left. \SI{4.0}{\per \s}.}
    \label{fig:h_profile}
\end{figure}

\begin{figure}[!ht]
\centering
\begin{adjustwidth}{-1cm}{0cm}
\begin{tabular}{l l}
\includegraphics[width=0.50\textwidth]{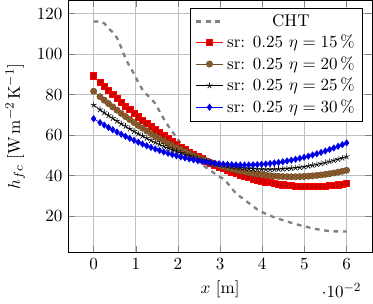} & \includegraphics[width=0.50\textwidth]{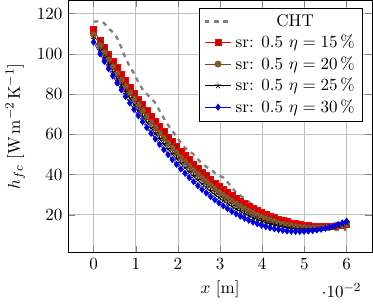} \\ 
\includegraphics[width=0.50\textwidth]{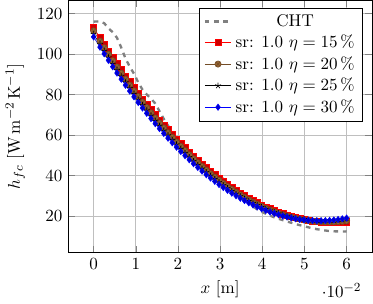} & \includegraphics[width=0.50\textwidth]{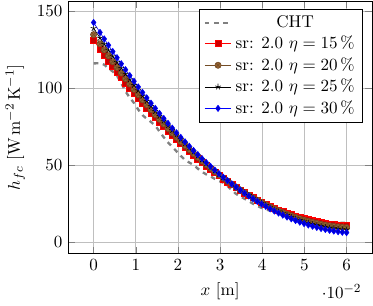}\\
\includegraphics[width=0.50\textwidth]{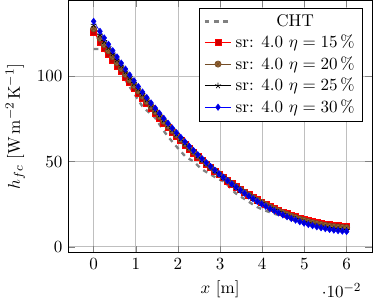} & \\
\end{tabular}
\end{adjustwidth}
\caption{Estimated profiles of the convective heat transfer coefficient on the upper surface of the disc $h_fc$ at different sampling rates for additive Gaussian noise range \SIrange{15}{30}{\percent}.  Top Left. \SI{0.25}{\per \s} . Top Right. \SI{0.50}{\per \s}. Middle left. \SI{1.0}{\per \s}. Middle right. \SI{2.0}{\per \s}. Bottom left. \SI{4.0}{\per \s}.}
\label{fig:h_profile_high}
\end{figure}

\Cref{fig:h_profile,fig:h_profile_high} show the comparison between \gls{pinn}-based estimation with \gls{cht}-based estimation of \gls{chtc} profile. At noise levels typical of experimental setups (up to \SI{10}{\percent}) and with a sampling rate of \SI{0.5}{\per \second} or greater, the estimated $h_{fc}(x)$ shows strong qualitative agreement with the \gls{cht} benchmark(see \Cref{fig:h_profile}). Notably, the \gls{pinn}-based estimation effectively captures the prominent peak near the stagnation region and the subsequent decline as one moves radially away from the jet center.

However, as the noise level increases—particularly beyond \SI{15}{\percent} or higher, low-resolution measurements become less reliable, leading to noticeable deviations of the $h_{fc}(x)$ profile (see \Cref{fig:h_profile_high}). In contrast, data with higher sampling rates demonstrate significantly greater robustness under high-noise conditions (e.g. \SIrange{15}{30}{\percent}), successfully preserving sharp gradients and local variations in the estimated $h_{fc}(x)$. This suggests that while \gls{pinn}-based estimation performs well in low-noise scenarios even with low-frequency data, high-resolution sampling is essential for maintaining accuracy and stability when noise levels are elevated.

\begin{figure}[!htp]
    \centering
\begin{adjustwidth}{-3cm}{0cm}
\begin{tabular}{l l}
\includegraphics[width=0.75\textwidth]{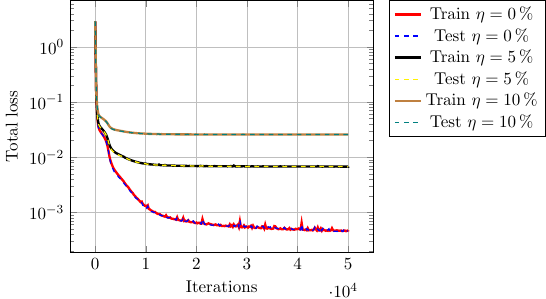} & \includegraphics[width=0.50\textwidth]{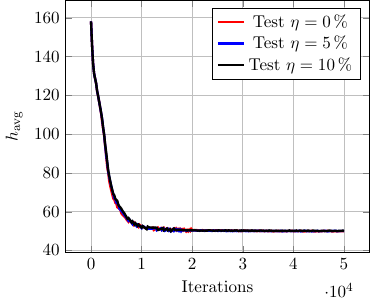} \\ 
\includegraphics[width=0.75\textwidth]{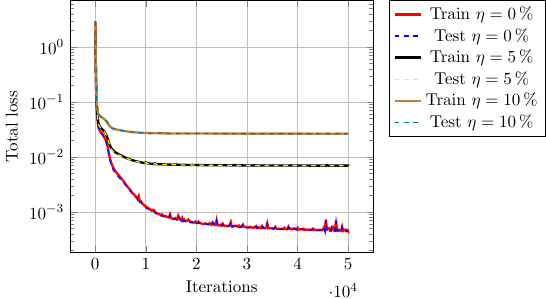} & \includegraphics[width=0.50\textwidth]{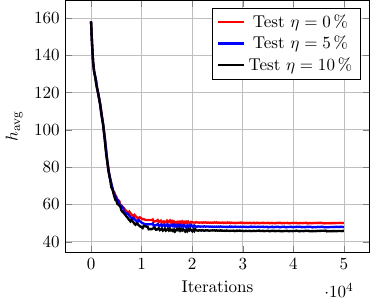} \\
\includegraphics[width=0.75\textwidth]{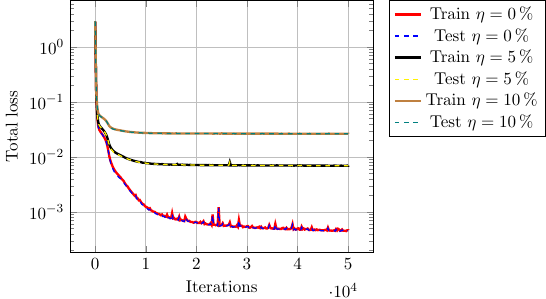} & \includegraphics[width=0.50\textwidth]{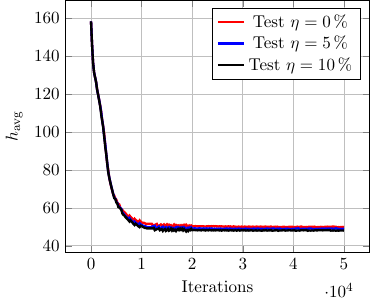}
\end{tabular}
\end{adjustwidth}
\caption{Convergence plots of total loss for both training and test data (left) and averaged value of estimated $h_{fc}(x)$  obtained via numerical integration (right) for additive Gaussian noise \SIlist{0; 5; 10}{\percent} for different sampling rates.  Top. \SI{0.25}{\per \s}. Middle \SI{0.50}{\per \s}. Bottom. \SI{1.0}{\per \s}.}
    \label{fig:h_loss_conv_func1}
\end{figure}

\begin{figure}[!ht]
    \centering
    \begin{adjustwidth}{-3cm}{0cm}
 \begin{tabular}{l l}
\includegraphics[width=0.75\textwidth]{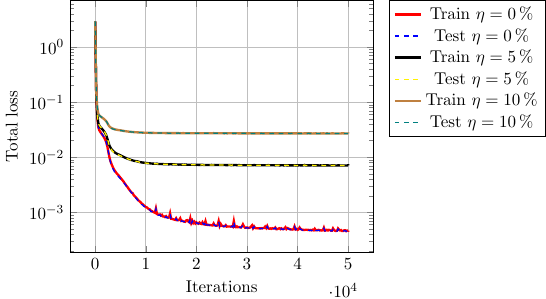} & \includegraphics[width=0.50\textwidth]{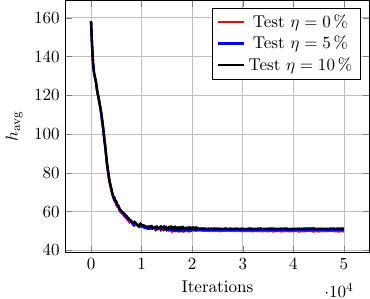} \\
\includegraphics[width=0.75\textwidth]{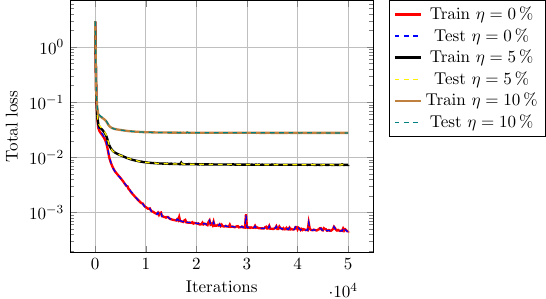} & \includegraphics[width=0.50\textwidth]{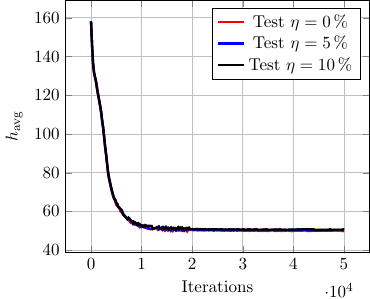}
\end{tabular}
 \end{adjustwidth}
\caption{Convergence plots of total loss for both training and test data (left) and averaged value of estimated $h_{fc}(x)$ obtained via numerical integration (right) for additive Gaussian noise \SIlist{0; 5; 10}{\percent} for different sampling rates.  Top. \SI{2,0}{\per \s}. Bottom. \SI{4.0}{\per \s}.}
\label{fig:h_loss_conv_func2}
\end{figure}

\Cref{fig:h_loss_conv_func1,fig:h_loss_conv_func2} presents the convergence plots of the total loss function and the spatially averaged value of the estimated $h_{fc}(x)$, denoted as $h_{\text{avg}}$, obtained by numerically integrating the learned function over the impingement surface. The figures illustrate results for three noise levels: \SIlist{0;5;10}{\percent} across all five sampling rates: \SIlist{0.25; 0.5; 1.0; 2.0; 4.0}{\per \second}. 
The left panel shows the decay of the loss function over training iterations, while the right panel depicts the convergence behavior of the averaged estimate $h_{\text{avg}}$.


To assess the robustness of the framework under severe measurement uncertainty, additional simulations were performed with noise levels up to \SI{30}{\percent}. As shown in \Cref{tab:noise_sr_error_avg_8_sensors}, the results indicate that, similar to the constant-coefficient case, higher sampling rates become increasingly important at elevated noise levels. However, under low-noise conditions, coarser temporal data still yield satisfactory performance.

\begin{table}[!htp]
    \centering
    \begin{tabular}{c|rrrrr}
\toprule
\diagbox[width=4.5cm]{Noise (\si{\percent})}{Sampling rate}& 0.25 & 0.50 & 1.00 & 2.00 & 4.00 \\
\midrule
0.0 & 6.395 & 6.253 & 6.139 & 6.110 & 6.261 \\
2.0 & 9.109 & 5.643 & 6.186 & 5.901 & 6.179 \\
4.0 & 12.188 & 5.672 & 6.264 & 5.680 & 5.778 \\
5.0 & 14.063 & 5.794 & 6.241 & 5.592 & 5.721 \\
6.0 & 15.688 & 5.904 & 6.543 & 5.998 & 5.628 \\
8.0 & 19.086 & 6.743 & 6.798 & 6.393 & 5.640 \\
10.0 & 22.381 & 7.624 & 7.032 & 6.804 & 5.529 \\
15.0 & 29.914 & 10.601 & 8.358 & 8.738 & 6.266 \\
20.0 & 36.959 & 14.313 & 9.627 & 11.126 & 7.160 \\
25.0 & 43.684 & 17.915 & 11.165 & 13.895 & 8.147 \\
30.0 & 50.376 & 21.428 & 12.467 & 16.480 & 9.387 \\
\bottomrule
\end{tabular}
\caption{Relative error (defined in \Cref{eqn:error_profile})  in the estimated convective heat transfer profile using \glspl{pinn} taking with convective heat transfer profile obtained from \gls{cht} simulations as baseline, for various levels of additive Gaussian noise and sampling rate.}
    \label{tab:noise_time_error}
    \end{table}

The relationship between noise and estimation error remains non-monotonic: in some settings, moderate noise yields slightly improved generalization due to implicit regularization during loss optimization. This aligns with our observations in the previous test case.

Overall, these results underscore the robustness of the PINNs framework, which maintains reliable performance even under noisy and sparsely sampled conditions—a setting where traditional inverse methods are known to degrade significantly in accuracy and stability \cite{engl_regularization_1996}.
\section{Conclusions}\label{sec:conclusion}
This study presented a \glsxtrfull{pinn} framework for the inverse estimation of \glsxtrfull{chtc} in a jet impingement cooling scenario. Using synthetic temperature measurements derived from high-fidelity \glsxtrfull{cht} simulations, we have demonstrated that the proposed method accurately recovers both averaged and spatially varying CHTC at the fluid–solid interface. 

The constant-coefficient case served as a baseline to test the model’s numerical stability and sensitivity to noise, while the spatially varying case captured realistic, non-uniform thermal boundary behavior. Across both cases, the \glspl{pinn} model exhibited strong agreement with ground truth values, maintaining robustness even under significant synthetic measurement noise and limited temporal resolution. The model’s performance is evaluated up to \SI{30}{\percent} of additive Gaussian noise and five temporal sampling rates \SIlist{0.25; 0.5; 1.0; 2.0; 4.0}{\per \second}. For noise levels up to \SI{10}{\percent}  and a sampling rate of \SI{0.5}{\per \second} or higher, the predicted spatially varying \gls{chtc} profile $h_{fc}(x)$ shows strong agreement (within \SI{8}{\percent}) with \gls{cht}-derived benchmarks. Even under high-noise conditions, the \gls{pinn} framework was able to predict the \gls{chtc} profile within a practically acceptable error margin when provided with densely sampled data. Notably, the framework does not require explicit modeling of the fluid domain, making it particularly attractive for practical applications where flow conditions are unknown or difficult to model.

These results suggest that \glspl{pinn} offer a viable and flexible alternative to traditional inverse heat transfer methods and CHT-based inference, particularly in scenarios with sparse, uncertain, or noisy data. Future work will focus on extending this framework to real experimental datasets, optimal placement of sensors, estimation of a spatial-temporally varying \gls{chtc}, incorporating uncertainty quantification, and applying the methodology to more complex cooling configurations such as multi-jet arrays or rotating systems.    

\clearpage
\section*{Appendix A: Grid independence study} 
\label{app:gridind}
To ensure that the \glsxtrshort{cht} simulations used as the ground truth for inverse estimation are not significantly affected by spatial discretization errors, a grid independence study was conducted. This study evaluates the sensitivity of the \glsxtrshort{chtc}—to mesh refinement near the fluid–solid interface and in the jet impingement region.

\begin{figure}[!ht]
	\centering
	\includegraphics[width=0.9\textwidth]{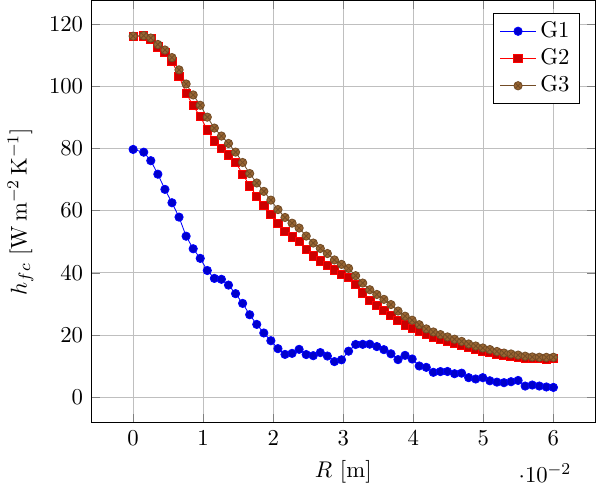}
	\caption{Grid independence study showing the variation of spatial distribution of the convective heat transfer coefficient over the impingement surface $h_{fc}$ for three different mesh resolutions. The negligible difference between medium and fine grids indicates numerical convergence, validating the choice of the medium grid for subsequent simulations.}
	\label{fig:grid_ind}
\end{figure}

Three structured hexahedral meshes of increasing resolution were considered:

\begin{itemize}
    \item \textbf{Coarse Grid (G1)}: \num{450602} elements
    \item \textbf{Medium Grid (G2)}: \num{3245782} elements
    \item \textbf{Fine Grid (G3)}: \num{5412698} elements
\end{itemize}
The resulting \glsxtrshort{chtc} distributions along the impingement surface were plotted as a function of radial position in \Cref{fig:grid_ind}.

A visual comparison of these profiles shows excellent agreement between the medium and fine grids, with nearly overlapping curves across the entire domain. This close match suggests that further mesh refinement does not meaningfully alter the predicted heat transfer behavior. Based on this observation, the medium-resolution grid (G2) was selected for all subsequent \glsxtrshort{cht} simulations, offering a practical trade-off between computational efficiency and solution fidelity. This visual verification reinforces the consistency and reliability of the CFD data used for benchmarking the \glspl{pinn}-based inverse estimation.
\section*{Acknowledgements}
Arijit Hazra would like to acknowledge funding support from the Ramanujan Fellowship (RJF/2022/000046) administered by SERB-DST (Now ANRF), India.

\end{document}